\documentclass[10pt,letterpaper]{article}
\usepackage{opex3}
\usepackage{amsmath,amssymb}        
\usepackage[mathscr]{euscript}      
\usepackage{dsfont}                 
\usepackage{graphicx}

\usepackage{color} 

\begin{document}


\title{All-reflective coupling of two optical cavities with 3-port diffraction gratings}

\author{Oliver Burmeister$^1$, Michael Britzger$^1$, Andr\'{e} Th\"uring$^1$, Daniel Friedrich$^1$,  Frank Br\"uckner$^2$, Karsten Danzmann$^1$, and Roman Schnabel$^1$}

\address{$^1$Albert-Einstein-Institut, Max-Planck-Institut f\"ur Gravitationsphysik and Leibniz Universit\"at Hannover, Callinstr. 38, 30167 Hannover, Germany\\
$^2$Institut f\"ur Angewandte Physik, Friedrich-Schiller-Universit\"at Jena, Max-Wien-Platz 1, 07743 Jena, Germany}

\email{roman.schnabel@aei.mpg.de}


\begin{abstract}
The shot-noise limited sensitivity of Michelson-type laser interferometers with Fabry-Perot arm cavities can be increased by the so-called power-recycling technique. In such a scheme the power-recycling cavity is optically coupled with the interferometer's arm cavities. A problem arises because the central coupling mirror transmits a rather high laser power and may show thermal lensing, thermo-refractive noise and photo-thermo-refractive noise. Cryogenic cooling of this mirror is also challenging, and thus thermal noise becomes a general problem. Here, we theoretically investigate an all-reflective coupling scheme of two optical cavities based on a 3-port diffraction grating. We show that power-recycling of a high-finesse arm cavity is possible without transmitting any laser power through a substrate material. The power splitting ratio of the three output ports of the grating is, surprisingly, noncritical. 
\end{abstract}



\section{Introduction}
\begin{figure}[htbp]
	\centering
		\includegraphics[height=4.1cm]{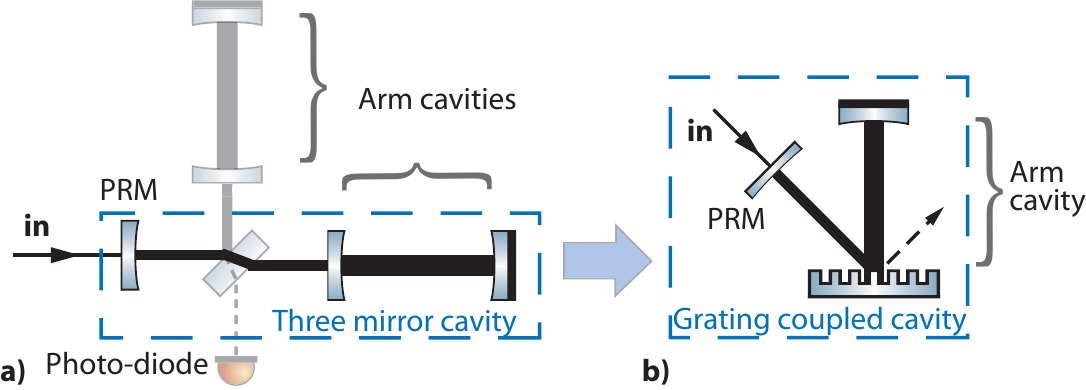}
	\caption{a) In laser-interferometric gravitational wave detectors two coupled cavities are formed by the two mirrors of an arm cavity and the power-recycling mirror (PRM). b) The transmissive coupling mirror can be replaced by an all-reflective 3-port grating. Without laser transmission thermal lensing and substrate thermal noise sources can be substantially suppressed.}
	\label{fig:Fig01}
\end{figure}

Coupled optical cavities have applications in 
optical-feedback laser devices \cite{DHD87,Wicht05} and in ultra-high-precision laser interferometers. Such cavities form a basis for proposed quantum-non-demolition interferometers \cite{BraginskyBar}, and were theoretically as well as experimentally analysed in order to reduce the quantum noise in displacement measurements \cite{meers}-\cite{TGVMDS09}. Meanwhile, coupled optical cavities are routinely used in today's laser-interferometric gravitational wave detectors \cite{LIGO2009,Virgo2006}. Here, their purpose is to resonantly enhance the laser power inside the interferometer arms without decreasing the detector's bandwidth too much. Fig.~\ref{fig:Fig01}(a) shows the topology used in operating gravitational wave detectors. The interferometer is operated close to a dark fringe and all the light that is back-reflected from the arm cavities is resonantly enhanced inside the so-called power-recycling cavity \cite{meers,recycling}. The power enhancement factor in this cavity also applies as an additional enhancement factor to the power inside the arm cavities, but, most importantly, without decreasing their signal bandwidth. The optimal power-recycling factor is realized if no light is reflected towards the laser source. In this case the coupled system of power-recycling cavity and arm cavity is impedance matched. The coupled cavities in this topology can be modelled by the three mirrors in the dashed box in Fig.~\ref{fig:Fig01}(a). The central mirror provides the coupling of the two cavities involved. Obviously, this component is exposed to the transmission of a rather high laser power. Since all materials show some residual optical absorption, the transmitted light heats the bulk material. Generally, the refractive index of a material depends on temperature which causes thermal lensing \cite{lensing} and photon absorption induced thermo-refractive noise \cite{BGV99,BGV00}. Residual optical absorption in mirrors also makes their cryogenic cooling challenging \cite{Cryo09}, and thermal noise \cite{Levin98} remains a general problem in high-precision laser interferometers with transmissive mirrors.

In order to circumvent these problems, all-reflective interferometer concepts have been developed where all (partially) transmitting mirrors are replaced by reflection gratings \cite{Drever}-\cite{gratingbs}. In a recent experiment a power-recycled Michelson interferometer with an dielectric all-reflective beamsplitter was demonstrated \cite{gratingbs}. 
We note, that the reflection gratings used in all these experiments incorporated dielectric multi-layer coating stacks, which also lead to light absorption and thermal noise. However, in principle multi-layer coating stacks can be avoided and their purpose be mimicked by appropriately designed surface nano-structures. Recently, we demonstrated a monolithic high-reflectivity nano-structured surface of a silicon crystal \cite{BruecknerPRL}. In such a scheme no (lossy) material is added to the substrate and surface absorption should not be an issue. The thermal noise of such a monolithic mirror is also expected to be lower than that of a multi-layer coating stack, however, a direct thermal noise comparison has not yet been performed. \\

Here, we theoretically analyze an all-reflective concept for the coupling of two optical cavities using a diffraction grating with three diffraction orders, i.e. with  three ports. As it turns out, in such a scheme the grating's diffraction efficiencies should be much smaller than its specular reflectivity. Consequently, only shallow grating structures are required which eases the grating fabrication process. If a 3-port diffraction grating is used as a coupling device, the arm cavity can be established perpendicular to the grating surface. The finesse of the arm cavity is then limited by the (high) specular reflectivity of the grating but not by its diffraction efficiency, see Fig.~\ref{fig:Fig01}(b). The coupling into the cavity is provided by the (small) first order diffraction efficiency of the grating. The second-order diffraction is aligned back towards the laser input which can be seen as a second-order Littrow configuration. We provide a detailled analysis of the arm cavity power build-ups in such an arrangement. Generally, the intra-cavity power build-up of grating coupled cavities depends on the cavity detuning, i.e. the microscopic length of the cavities, and on the diffraction efficiencies of each diffraction order of the grating. 
In particular, the existence of the third grating port [dashed arrow towards the right in Fig.~\ref{fig:Fig01}(b)] might be expected to influence the achievable power build-ups. 
Note that two such systems, as shown in Fig.~\ref{fig:Fig01}(b), can be combined to a gravitational wave antenna by inserting an (all-reflective) beam splitter between the power-recycling mirror and the grating(s). Gravitational wave signals can then photo-electrically be detected in the (almost) dark port of the beam splitter, similar to the topology in Fig.~\ref{fig:Fig01}\,(a). Signals also leave the arm cavities towards the open ports [dashed arrow in Fig.~\ref{fig:Fig01}(b)], and can be detected with two balanced homodyne detectors, one for each arm. If the open ports also contain carrier light, the two fields in the open ports may be overlapped on another balanced beam splitter and the output field in the (almost) dark port again can be detected by a single photo-diode. If the signals in all ports are detected properly and finally added, the same signal-to-shot-noise ratio is achieved as in the conventional topology shown in Fig.~\ref{fig:Fig01}(a).

\section{Two cavities coupled by a conventional mirror} 
\label{sec:ConventionalCoupling}
Throughout this investigation we consider all optical components to be loss-less. In this case a conventional, partially transmitting mirror can be fully  characterized by a $2\times 2$ scattering matrix acting on two-dimensional vectors whose components represent the amplitudes of the input fields. One possible representation of this matrix is 
\begin{equation}
\mathbf{S}=
	\begin{pmatrix}
		\rho			&\mathrm{i}\tau\\
		\mathrm{i}\tau	&\rho
	\end{pmatrix}\label{3p/strmatr/eq:7},
\end{equation}
where $\rho$ and $\tau$ are the mirror's amplitude reflectivity and transmittance, respectively. The complex $\mathrm{i}$ accounts for the phase gained in transmission of the mirror. Energy conservation further implies
\begin{equation}
	\rho^2+\tau^2=1\,.
\end{equation}

\begin{figure}[htbp]
	\centering
		\includegraphics[height=3.1cm]{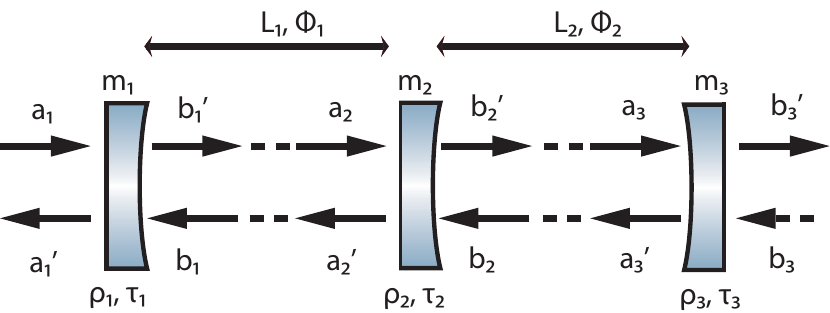}
	\caption{Labeling of the optical fields at a linear three-mirror cavity. Every mirror $\mathrm{m}_i$ has two ports with pairs of in- and output field amplitudes denoted by the vectors ($\mathrm{a}_i, \mathrm{b}_i$) and ($\mathrm{a}_i', \mathrm{b}_i'$), respectively, and the amplitude transmission and reflection coefficients $\rho_i$ and $\tau_i$. Thus two optically coupled cavities with the macroscopic length $\mathrm{L}_i$ and the microscopic detuning parameter $\Phi_i$ are formed. For most considerations in this paper we choose $\rho_1^2=0.7$, $\rho_2^2=0.8$, $\rho_3^2=0.9$, and $\mathrm{b}_3=0$. For possible applications in high-power laser interferometry we also discuss parameter sets with $\rho_3$-values close to unity.}
	\label{fig:Fig02}
\end{figure}

Figure~\ref{fig:Fig02} resembles the two coupled cavities of Fig.~\ref{fig:Fig01}. In accordance to the application of such an arrangement in a gravitational-wave detector the first cavity (composed of the mirrors $\mathrm{m}_1$ and $\mathrm{m}_2$) shall be denoted as `recycling cavity', while the second cavity (composed of the mirrors $\mathrm{m}_2$ and $\mathrm{m}_3$) shall be called `arm cavity'. In Fig.~\ref{fig:Fig02},   $\mathrm{a}_i$ and $\mathrm{b}_i$ denote the amplitudes of the optical fields propagating towards the mirror m$_i$ and $\mathrm{a}_i'$ and $\mathrm{b}_i'$ correspond to the fields propagating away from the mirror.  It is assumed that the incident light fields match the fundamental spatial mode of the cavity set-up. The length of each cavity can be written as $\mathrm{L}_i=\lambda(\mathrm{l}_i+\Phi_i/(2\pi)) $, where $\mathrm{l}_i$ is an integer denoting the macroscopic length of the cavity in multiples of the wavelength $\lambda$, and the detuning parameter $\Phi_i$ is restricted to the range $-\pi<\Phi_i<\pi$ giving the microscopic length as a fractional part of the wavelength. Throughout this investigation a single optical frequency is considered and the guoy phase is omitted. Thus the phase change associated with one round trip along the cavity is accounted for by multiplying the relevant field by the factor $\exp{(2\mathrm{i}\Phi_i)}$.   

In order to calculate the power build-up in the second cavity the field amplitude b$_3'$ needs to be derived. It is convenient to start with the fields at the second cavity that, for b$_3=0$, can be written as: 
\begin{eqnarray}
a_2'&=&\mathrm{i}\tau_2 b_2+\rho_2 a_2,\\ 
	b_2'&=&\mathrm{i} \tau_2 a_2+\rho_2  b_2,\label{eq:b2'}\\
	b_2 &=& \rho_3 e^{2\mathrm{i}\Phi_2}b_2'\label{eq:b2}.
\end{eqnarray}
Inserting Eq.~(\ref{eq:b2}) into Eq.~(\ref{eq:b2'}) yields   
\begin{equation}
	b_2'=\frac{ \mathrm{i} \tau_2	}{1-\rho_2 \rho_3 e^{2\mathrm{i}\Phi_2}}a_2\label{eq:b2strich}.
\end{equation}
This expression is equivalent to the well-known case of a two-mirror cavity, however a fundamental difference lies in the input field $a_2$, which is not fixed but depends on the detuning of both cavities $\Phi_1$ and $\Phi_2$. 
It can also be shown that 
\begin{equation}
	a_2'=\frac{\rho_2-\rho_3 e^{2\mathrm{i}\Phi_2}}{1-\rho_2 \rho_3 e^{2\mathrm{i}\Phi_2}} a_2\,.
	\label{eq:a_2'}
\end{equation}
The complex coefficient in Eq.~(\ref{eq:a_2'}) is the complex reflection coefficient of the second cavity as seen from mirror $\mathrm{m}_1$. In this sense the second cavity can be interpreted as a \textit{compound} mirror with reflectivity
\begin{equation}
	\rho_{\mathrm{m}_2\mathrm{m}_3}=\frac{\rho_2-\rho_3 e^{2\mathrm{i}\Phi_2}}{1-\rho_2 \rho_3 e^{2\mathrm{i}\Phi_2}}\,.
	\label{eq:cmr_rm2m3}
\end{equation}
Note, that the complex reflectivity not only depends on mirror reflectivities but also on the detuning of the second cavity ($\Phi_2$). The phase shift that is gained due to the reflection of the field at the second cavity can be calculated according to $\mathrm{arg}[\rho_{\mathrm{m}_2\mathrm{m}_3}(\Phi_2)]$. 
 
The field inside the first cavity is then given by the equation for a simple two-mirror cavity: 
\begin{equation}
	b_1'=\frac{\mathrm{i}\tau_1}{1-\rho_1 \rho_{\mathrm{m}_2\mathrm{m}_3}e^{2\mathrm{i}\Phi_1}} a_1. 
	\label{eq:b1'}
\end{equation}
And the fields that are either reflected at or transmitted through the three-mirror cavity, respectively, can be written as:
\begin{eqnarray}
	a_1'&=&\frac{\rho_1-\rho_{\mathrm{m}_2\mathrm{m}_3}e^{2\mathrm{i}\Phi_1 }}{1-\rho_1\rho_{\mathrm{m}_2\mathrm{m}_3}e^{2\mathrm{i}\Phi_1}}a_1,\label{eq:rho_3MC}\\
	b_3'&=&-\frac{\tau_2\tau_3 e^{\mathrm{i}\Phi_2}}{1-\rho_2\rho_3 e^{2 \mathrm{i}\Phi_2}}{b_1'}\cdot e^{\mathrm{i}\Phi_1}. 
	\label{eq:tau_3MC}
\end{eqnarray}

\begin{figure}[htbp]
	\centering
		\includegraphics[height=4.05cm]{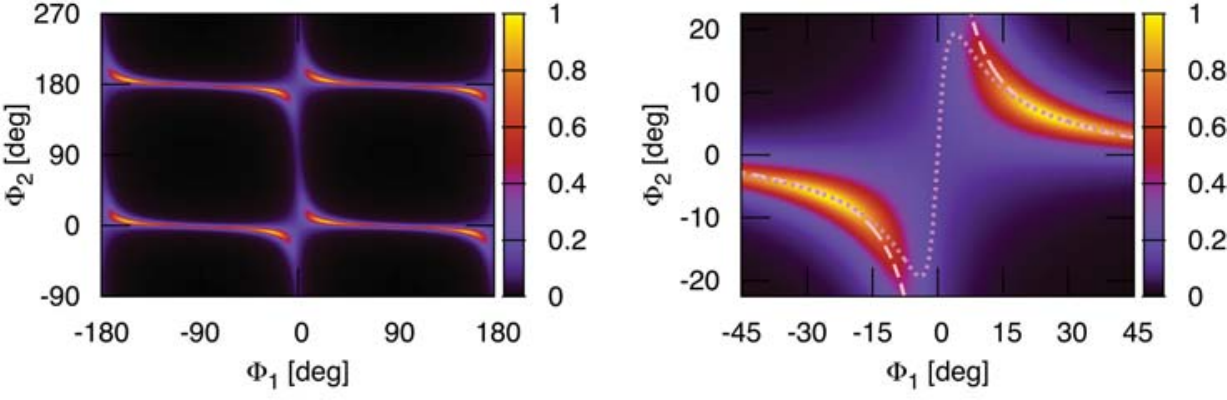}
	\caption{Transmitted power at a loss-less three-mirror cavity with $\rho_1^2=0.7, \rho_2^2=0.8$ and $\rho_3^2=0.9$ as a function of the detunings $\Phi_1$ and $\Phi_2$. Here, on peak resonance all power is transmitted. The lines in the zoomed-in figure on the right represent the resonance branches of the individual cavities given by Eqs.~(\ref{eq:phi1_res}) (dashed) and (\ref{eq:phi2_res}) (dotted), respectively.}
	\label{fig:Fig03}
\end{figure}

Field $b_3'$ in Eq.~(\ref{eq:tau_3MC}) is proportional to the power stored in the second cavity. Together with Eq.~(\ref{eq:b1'}) an equation is found  that relates the input field $a_1$ to the transmitted field $b_3'$. For a steady state of the coupled cavities the maximum of the absolute value of the latter is identical to the absolut value of the input field. In this case the coupled cavities are impedance matched. In Fig.~\ref{fig:Fig03} we illustrate the resonance behaviour of the three-mirror cavity as shown in Fig.~\ref{fig:Fig02} with mirror reflectivities chosen to be $\rho_1^2=0.7,\rho_2^2=0.8$ and $\rho_3^2=0.9$. The plot shows the input normalized transmitted power ($|b'_3|^2/|a_1|^2$) as a function of the cavity detunings $\Phi_1$ and $\Phi_2$ (color coded). The resonance pattern is periodic in $\Phi_1\,\mathrm{mod}\,\pi$ and in $\Phi_2\,\mathrm{mod}\,\pi$ and point symmetric with respect to the origin. For some combinations of detunings impedance matching is achieved. In this case the total power build-up inside the second cavity is simply given by $1/(1-\rho_3^2)$.

The resonance pattern of the three-mirror cavity as shown in Fig.~\ref{fig:Fig03} can be understood by calculating the resonances of the individual cavities. A two-mirror cavity is on resonance if the accumulated phase delay per round trip is zero, i.e. an integer multiple of $2\pi$. Consequently, the resonance of the first cavity of the coupled system is achieved if the microscopic position of $\mathrm{m}_1$ is chosen such that it compensates half the phase shift due to the reflection of the field at the second cavity:
\begin{equation}
	\Phi_1^{\mathrm{res}}=-\frac{1}{2} \mathrm{arg}[ \rho_{\mathrm{m}_2\mathrm{m}_3}(\Phi_2)]\,.
	\label{eq:phi1_res}
\end{equation}

One can also deduce an expression for the complex compound mirror reflectivity of the first cavity as seen from mirror $\mathrm{m}_3$:
\begin{equation}
	\rho_{\mathrm{m}_2\mathrm{m}_1}=\frac{\rho_2-\rho_1 e^{2\mathrm{i}\Phi_1}}{1-\rho_1 \rho_2 e^{2\mathrm{i}\Phi_1}}\,,
	\label{eq:cmr_rm2m1}
\end{equation}
where $\Phi_1$ is the detuning of the first cavity. Accordingly, the resonance of the second cavity is given for the following detuning:
\begin{equation}
	\Phi_2^{\mathrm{res}}=-\frac{1}{2} \mathrm{arg}[ \rho_{\mathrm{m}_2\mathrm{m}_1}(\Phi_1)]\,.
	\label{eq:phi2_res}
\end{equation}

The resonance branches of the individual cavities are given as dashed [Eq.~(\ref{eq:phi1_res})] and dotted lines [Eq.~(\ref{eq:phi2_res})] in Fig.~\ref{fig:Fig03}. Obviously, maximum transmitted power is achieved if both individual cavities are on resonance, i.e. in the crossing points of the resonance branches. 

\begin{figure}[htbp]
	\centering
		\includegraphics[height=4.56cm]{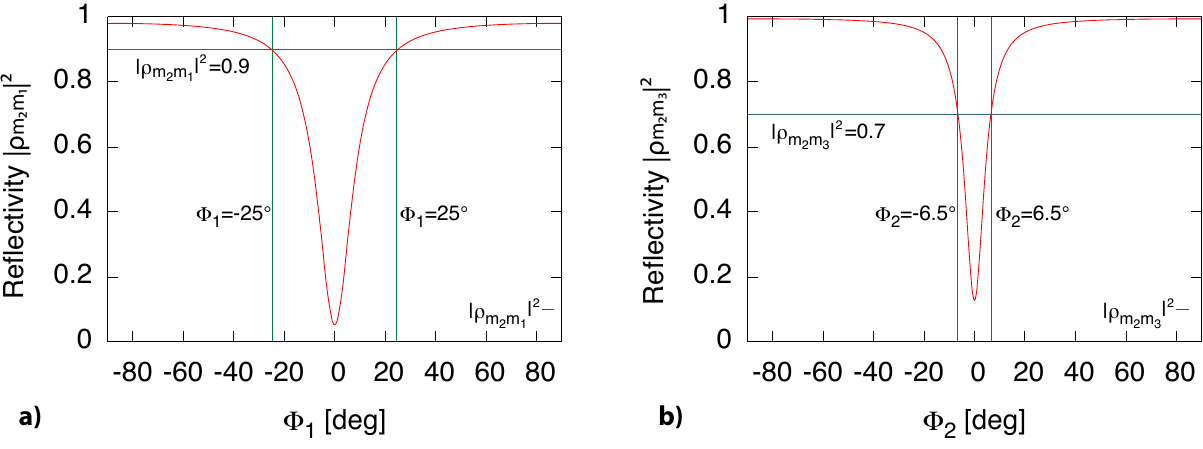}
	\caption{Compound mirror reflectivities $|\rho_{\mathrm{m}_2\mathrm{m}_1}|^2$ (a) and $|\rho_{\mathrm{m}_2\mathrm{m}_3}|^2$ (b) as a function of $\Phi_1$ and $\Phi_2$, respectively. The maximum of the transmitted power occurs at the set of detunings where the compound mirrors matches the reflectivity of the respective single mirror $|\rho_{\mathrm{m}_2\mathrm{m}_1}(\Phi_1=\pm25^\circ)|^2={\rho_3}^2=0.7$ and $|\rho_{\mathrm{m}_2\mathrm{m}_3}(\Phi_2=\pm6.5^\circ)|^2={\rho_1}^2=0.9$.}
\label{fig:Fig04}
\end{figure}

Figure~\ref{fig:Fig04} shows the compound mirror reflectivities according to Eqs.~(\ref{eq:a_2'}) and (\ref{eq:cmr_rm2m1}) as a function of the detunings $\Phi_2$ and $\Phi_1$, respectively.   
Maxima of the transmitted power in Figure~\ref{fig:Fig03} are located at $\Phi_1=-25^\circ$ and $\Phi_2=-6.5^\circ$ and at $\Phi_1=25^\circ$ and $\Phi_2=6.5^\circ$. For this set of detunings the compound mirror reflectivities of the first and the second cavity fulfill the impedance matching condition
\begin{eqnarray}
|\rho_{\mathrm{m}_2\mathrm{m}_3}(\Phi_2)| &=& \rho_1 \,,\\
|\rho_{\mathrm{m}_2\mathrm{m}_1}(\Phi_1)| &=& \rho_3 \,,
\end{eqnarray}    
where no power is reflected but all light transmitted through the three-mirror cavity.

Let us now consider another set of reflectivities which is more related to an application in high-power laser interferometers such as gravitational-wave detectors. In order to achieve a maximum power build-up in the second cavity (the arm cavity) the reflectivity of the end mirror $\mathrm{m}_3$ is typically close to unity, e.g.  $\rho_3^2=0.99995$. The reflectivity of $\mathrm{m}_2$ is chosen in order to obtain a certain desired signal bandwidth of the arm cavity, e.g. $\rho_2^2=0.95$. In order to obtain impedance matching the reflectivity of $\mathrm{m}_1$ needs to be matched to $|\rho_{\mathrm{m}_2\mathrm{m}_3}|^2$ at the operating point of the arm cavity \cite{LIGO2009}. In analogy to the considerations above one finds that impedance matching is achieved at $\rho_1^2=0.996$ (at the operating point $\Phi_1=90^\circ$ and $\Phi_2=0^\circ$). One can show that in our example the power enhancement factor provided by the first cavity, i.e. by the power-recycling cavity, is 256, and the resonant power enhancement factor of the arm cavity is 78. The total power build-up in the arm cavity therefore is about 20,000. Again we have considered a situation without absorption and scattering. We note that in the presence of loss inside the first cavity, for example caused by the beam splitter of the Michelson interferometer (Fig.~\ref{fig:Fig01}), the reflectivity $\rho_1^2$ has to be reduced in order to achieve impedance matching.

\section{3-port grating coupler} 
\label{sec:three_port_grating_coupler}
\begin{figure}[htbp]
	\centering
		\includegraphics[height=2.306cm]{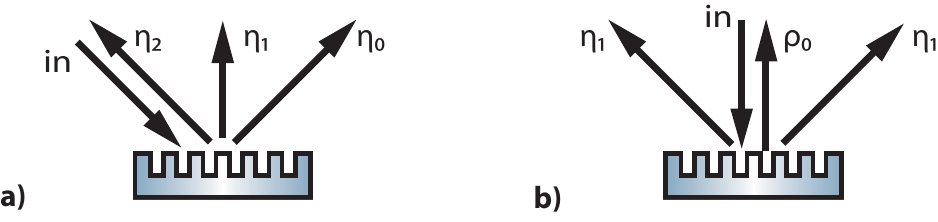}
	\caption{3-port grating coupler: (a) Labeling of amplitude diffraction coefficients for second-order Littrow configuration. (b) Amplitude diffraction coefficients and amplitude reflectivity for normal incidence.}
	\label{fig:Fig05}
\end{figure}

A loss-less optical component that splits an incoming field into three parts is described by a $3\times3$ scattering matrix. Here, we consider a 3-port diffraction grating in the 2nd oder Littrow arangement as investigated in Refs.~\cite{bunkowski2,bunkowski1}, see Fig.~\ref{fig:Fig05}(a). In such an arrangement the input beam under an angle is diffracted such that the 2nd order is retro-reflected towards the input beam and the 1st order is oriented along the gratings normal. The first order diffration can therefore provide the coupling into a Fabry-Perot cavity, see Fig.~\ref{fig:Fig01}(b). For normal incidence only 0th and $\pm$1st order diffraction are present, see Fig.~\ref{fig:Fig05}(b). 

Energy conservation yields the following relations for the amplitude coupling coefficients $\rho_0$, $\eta_0$, $\eta_1$, $\eta_2$:
\begin{eqnarray}
\rho_0^2+2\eta_1^2 &=&1 ,\\
\eta_0^2+\eta_1^2+\eta_2^2 &=&1.
\end{eqnarray}
As shown in Ref.~\cite{bunkowski1} the values for $\eta_0$ and $\eta_2$ have to be within the following boundaries
\begin{equation}
	\eta_{0\mathrm{_{min}^{max}}}=\eta_{2\mathrm{_{min}^{max}}}=\frac{1\pm\rho_0}{2}\,.
	\label{eq:etaminmax}
\end{equation}
The phase shifts associated with the coupling coefficients ($\phi_0$, $\phi_1$ and $\phi_2$) are generally functions of the diffraction efficiencies and can be found in Ref.~\cite{bunkowski1}. 

\section{Two cavities coupled by a 3-port grating} 
\label{sec:3PortCoupling}
 
\begin{figure}[htbp]
	\centering
		\includegraphics[height=6.64cm]{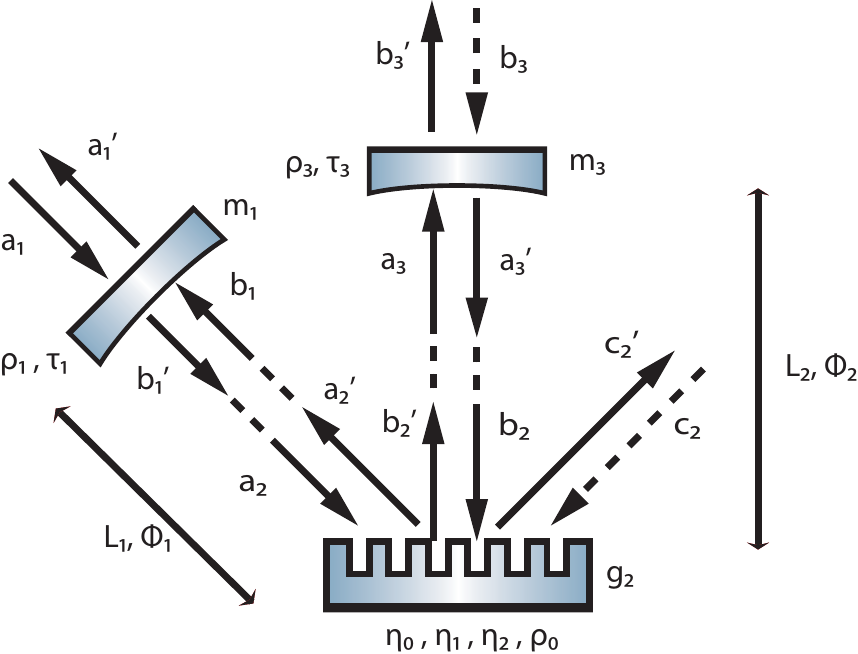}
	\caption{Schematic for an all-reflective coupling of two optical cavities by a 3-port grating. The amplitudes of the fields impinging on the optical component with subscript $i$ are denoted as a$_i$, b$_i$, and c$_i$. The output fields are denoted as a$'_i$, b$'_i$, and c$'_i$. The grating g$_2$ and mirror m$_1$ form the so-called power-recycling cavity of macroscopic length $\mathrm{L}_1$ and microscopic detuning parameter $\Phi_1$. The grating and mirror m$_3$ form the so-called arm cavity of macroscopic length $\mathrm{L}_2$ and microscopic detuning parameter $\Phi_2$.} 
	\label{fig:Fig06}
\end{figure}

In this Section we investigate the all-reflective coupling of two Fabry-Perot cavities with a 3-port reflection grating, as shown in Fig.~\ref{fig:Fig06}. We answer the question if such a scheme can reach a laser power enhancement factor and arm cavity bandwidth that is comparable with the conventional reflectivly coupled scheme described in Sec.~\ref{sec:ConventionalCoupling}. For the all-reflective coupling, mirror m$_2$ is replaced by a 3-port grating g$_2$. We start with a general analysis, and proceed with a consideration of three different grating designs that reflect the boundaries in Eq.~(\ref{eq:etaminmax}), namely the cases $\eta_{2\mathrm{min}}$, $\eta_{2\mathrm{max}}$ and $\eta_2=\eta_0=\eta_{0/2\mathrm{mid}}$.

In analogy to the previous Section the light fields at the arm cavity coupler can be written as:   
\begin{eqnarray}
a_2' &=& \eta_2 e^{\mathrm{i}\phi_2}a_2+\eta_1 e^{\mathrm{i}\phi_1} b_2,\\
b_2' &=& \eta_1 e^{\mathrm{i}\phi_1} a_2 +\rho_0 b_2,\\
b_2 &=& \rho_3 e^{2\mathrm{i}\Phi_2} b_2'.
\end{eqnarray}
The carrier field inside the arm cavity can be derived as:
\begin{equation}
	b_2'=\frac{\eta_1 e^{\mathrm{i}\phi_1}}{1-\rho_0\rho_3 e^{2\mathrm{i}\Phi_2}} a_2.
\end{equation}
Again the internal fields of both cavities are not independent of each other, and the compound mirror reflectivity of the second cavity $\rho_{\mathrm{g}_2\mathrm{m}_3}$ can be derived from:
\begin{equation}
	a_2'=\underbrace{\left(\frac{\rho_3\eta_1^2 e^{2\mathrm{i}(\phi_1+\Phi_2)} }{1-\rho_0\rho_3 e^{2\mathrm{i}\Phi_2}}+\eta_2 e^{\mathrm{i}\phi_2}\right)}_{\rho_{\mathrm{g}_2\mathrm{m}_3}}a_2.
\end{equation}
By using the compound mirror reflectivity of the second cavity the carrier field inside the first cavity is given by: 
\begin{equation}
	b_1'=\frac{\mathrm{i}\tau_1}{1-\rho_1 \rho_{\mathrm{g}_2\mathrm{m}_3}e^{2\mathrm{i}\Phi_1}} a_1.
\end{equation}

One can also derive the compound mirror reflectivity of the first cavity consisting of $\mathrm{m}_1$ and $\mathrm{g}_2$ as seen from m$_3$:
\begin{equation}
\rho_{\mathrm{g}_2\mathrm{m}_1}=\frac{\rho_1 \eta_1^2 e^{2 \mathrm{i}(\phi_1+\Phi_1) }}{1-\rho_1 \eta_2 e^{\mathrm{i}(\phi_2+2\Phi_1) }}+\rho_0.
\end{equation} 

In contrast to the three-mirror cavity, the power-recycled 3-port grating cavity has \textit{three} output ports: the first port contains the \textit{back-reflected} field towards the laser input ($\mathrm{a}_1'$), the second port contains the field \textit{transmitted} through the arm-cavity ($\mathrm{b}_3'$), and the third contains the \textit{forward-reflected} component of the input field ($\mathrm{c}_2'$). The fields at these output ports are given by:
\begin{eqnarray}
	\label{eq:3p_a_1'} 
	\mathrm{a}_1'&=&\left(\rho_1-\frac{\tau_1^2\rho_{\mathrm{g}2\mathrm{m}3}e^	{2\mathrm{i}\Phi_1}}{1-\rho_1\rho_{\mathrm{g}2\mathrm{m}3}e^{2\mathrm{i}\Phi_1}}\right)\cdot \mathrm{a}_1,\\
	\mathrm{b}_3'&=&-\frac{\tau_1\tau_3\eta_1 e^{\mathrm{i}(\phi_1+\Phi_1+\Phi_2)} }{(1-\rho_0\rho_3 e^{2\mathrm{i}\Phi_2})(1-\rho_1\rho_{\mathrm{g}2\mathrm{m}3} e^{2\mathrm{i}\Phi_1})}\cdot \mathrm{a}_1,\label{eq:3p_b_3'}\\
\mathrm{c}_2'&=&\left[\left(\eta_0 e^{\mathrm{i}\Phi_1}+ \frac{\rho_3\eta_1^2 e^{\mathrm{i}(2\phi_1+\Phi_1+2\Phi_2)}}{1-\rho_0\rho_3 e^{2\mathrm{i}\Phi_2}} \right)\cdot\frac{\mathrm{i}\tau_1}{1-\rho_1\rho_{\mathrm{g}2\mathrm{m}3}e^{2\mathrm{i}\Phi_1}}\right]\cdot \mathrm{a}_1 \label{eq:3p_c_2'}.
\end{eqnarray}
In the following subsections we calculate the resonance behaviour of the coupled cavity illustrated in Fig.~\ref{fig:Fig06}. With respect to Eq.~(\ref{eq:etaminmax}) we regard three gratings with different balancing of $\eta_0$ and $\eta_2$ but the same $\rho_0$. Firstly we assume the lowest possible value of $\eta_2=\eta_{2 \mathrm{min}}=(1-\rho_0)/2$, secondly the highest possible value of $\eta_2=\eta_{2 \mathrm{max}}=(1+\rho_0)/2$ and thirdly a midvalue $\eta_{2 \mathrm{mid}}$ of the two diffraction efficencies  $\eta_0=\eta_2$. In all cases the bandwidth of the arm cavity is kept constant and identical to the conventional three-mirror cavity. We therefore always choose a specular grating reflectivity of $\rho_0^2=0.8$ and a mirror m$_3$ reflectivity of $\rho_3^2=0.9$. In order to allow for a direct comparison we also start with a mirror m$_1$ reflectivity of $\rho_1^2=0.7$.

\subsection{Case 1: $\eta_2=\eta_{2 \mathrm{min}}$} 
\label{sub:case_1_eta_2_2 rm_}

\begin{figure}[htbp]
	\centering
		\includegraphics[height=4.05cm]{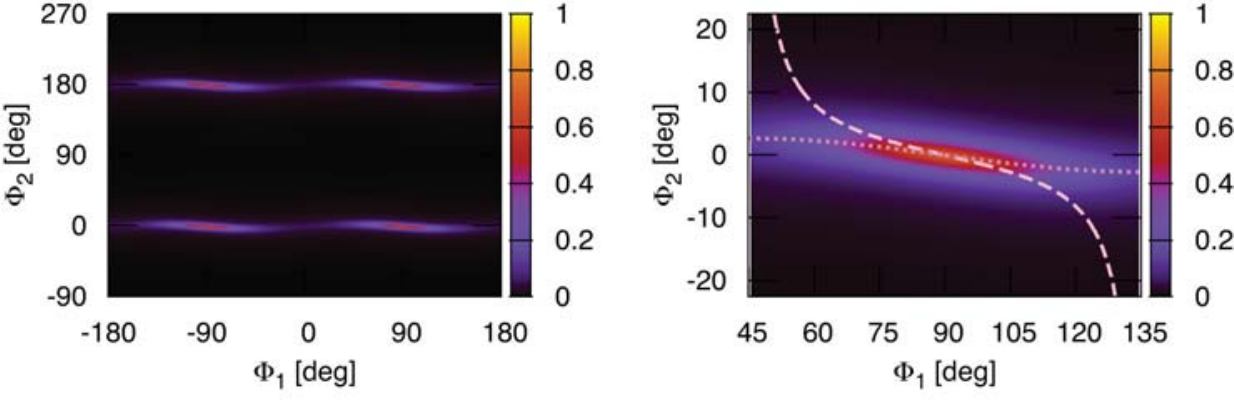}
	\caption{Transmitted power at a power-recycled 3-port cavity with $\rho_1^2=0.7, \rho_0^2=0.8$, $\eta_2=\eta_{2 \mathrm{min}}$ and $\rho_3^2=0.9$ as a function of the detunings $\Phi_1$ and $\Phi_2$. A moderate power enhancement is only present around a maximum at $\Phi_1=90^\circ$ and $\Phi_2=0^\circ$. The lines in the zoomed-in figure on the right represent the resonant detuning of the first cavity $\Phi_1^{\mathrm{res}}=-0.5 \mathrm{arg}[ \rho_{\mathrm{g}_2\mathrm{m}_3}(\Phi_2)]$ (dashed) to compensate the phase shift due to reflection at the second cavity and the resonant detuning of the second cavity $\Phi_2^{\mathrm{res}}=-0.5 \mathrm{arg}[ \rho_{\mathrm{g}_2\mathrm{m}_1}(\Phi_1)]$ (dotted), respectively.}
	\label{fig:Fig07}
\end{figure}

\begin{figure}[htbp]
	\centering
		\includegraphics[height=4.5cm]{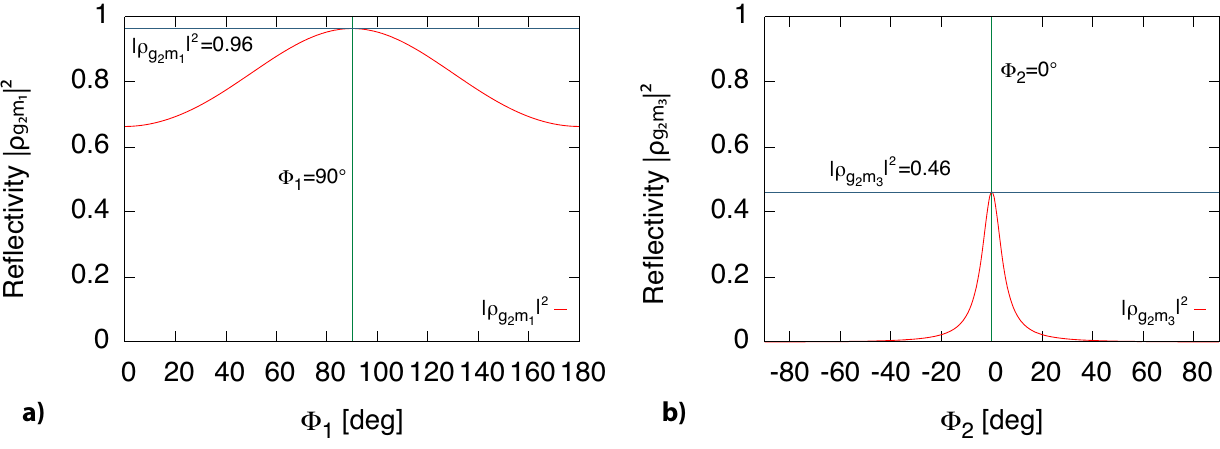}
	\caption{Compound mirror power reflectivities $|\rho_{\mathrm{g}_2\mathrm{m}_1}|^2$ (a) and $|\rho_{\mathrm{g}_2\mathrm{m}_3}|^2$ (b) for the $\eta_{2 \mathrm{min}}$-configuration as a function of $\Phi_1$ and $\Phi_2$, respectively. The resonance of the second cavity $\Phi_2=0^\circ$ corresponds to a \textit{maximum} of the compound mirror reflectivity of the cavity.}
	\label{fig:Fig08}
\end{figure}

Figure~\ref{fig:Fig07} shows the resonance pattern of a power-recycled 3-port grating cavity with $\rho_1^2=0.7$, $\rho_0^2=0.8$, $\rho_3^2=0.9$, and $\eta_2^2=\eta^2_{2 \mathrm{min}}=(1-\rho_0)^2/4 \approx 0.0028$. The normalized transmitted power ($|b'_3|^2/|a_1|^2$) is color-coded as a function of the detunings  $\Phi_1$ and $\Phi_2$. The resonance pattern is periodic in $\Phi_1\,\mathrm{mod}\,\pi$ and in $\Phi_2\,\mathrm{mod}\,\pi$. In contrast to the three-mirror cavity a significant power enhancement is only present around a maximum at $\Phi_1=90^\circ$ and $\Phi_2=0^\circ$. Even at this maximum only about 70\% of the light is transmitted through the cavity. The missing power is found in the retro-reflected and forward-reflected ports and the power build-up in the arm cavity of length L$_2$ is not optimal. 
Fig.~\ref{fig:Fig08}(b) shows why the retro-reflected power is not zero. The compound reflectivity of the arm cavity $|\rho_{\mathrm{g}_2\mathrm{m}_3}|^2$ has a maximal value of just 0.46. This value is significantly lower than the power-recycling mirror reflectivity of $\rho_1^2=0.7$. The power-recycling cavity is therefore not impedance-matched but under-coupled. Using Eq.~(\ref{eq:3p_a_1'}) one finds that on arm cavity resonance ($\Phi_2=0^\circ$) about 13\% of the power is retro-reflected. Impedance matching could be achieved with a decreased reflectivity of $\rho_1^2=0.46$. However, in this case the power transmitted through the arm cavity just increases from 70\% to 81\%. 
Using Eq.~(\ref{eq:3p_c_2'}) one finds that the still missing power of {19\%} is forward-reflected. The reason is that perfect destructive interference of this port can only be achieved for a perfect reflectivity of the arm cavity end mirror ($\rho_3^2=1$), see Ref.~\cite{bunkowski1}. \\
In order to maximize the power build-up in the arm cavities, $\rho_3^2$ should have values close to unity anyway. We therefore now consider again the application in a power-recycled interferometer and choose $\rho_0^2=0.95$ and $\rho_3^2=0.99995$. With these parameters the grating coupled arm cavity realizes a power enhancement factor of just 39, which is half the value of the conventional case. The reason is that for a 3-port grating coupler the arm cavity incoupling efficiency of $\eta_1^2=0.5\cdot(1-\rho_0^2) = 2.5 \%$ is half the incoupling in the conventional case ($\tau_2^2=1-\rho_2^2=5\%$). Consider now the additional power enhancement due to power-recycling. Following our previous discussion the largest power-recycling factor is achieved when the reflectivity of the power-recycling mirror meets $\rho_1^2=|\rho_{\mathrm{g}_2\mathrm{m}_3}(\Phi_2=0^\circ)|^2 \approx 0.998$. The corresponding transmission is only half the value for the power-recycling mirror in the conventional case. In turn, the power-recycling factor turns out to be 512 and is therefore twice as big as for the corresponding conventional case, see Sec.~\ref{sec:ConventionalCoupling}. The product of both build-up factors is about 20,000 and identical to the conventional scheme. The case considered here seems to be a reasonable all-reflective alternative to the conventional scheme in power-recycled interferometers. Note that almost no power is lost towards the forward-reflected port since in our last example $\rho_3$ was almost unity.


\subsection{Case 2: $\eta_2=\eta_{2 \mathrm{max}}$ } 
\label{sub:case_2_}
\begin{figure}[htbp]
 	\centering
 		\includegraphics[height=4.05cm]{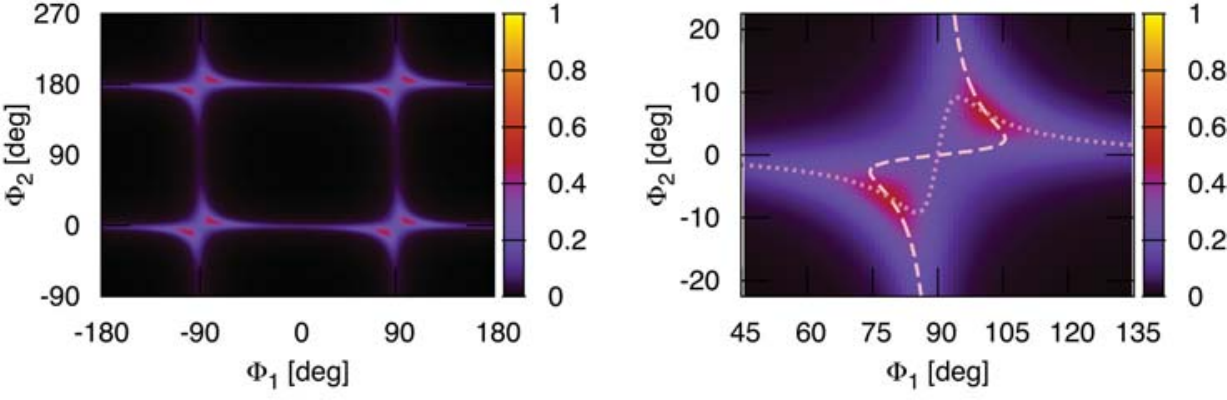}
 	\caption{Transmitted power at a power-recycled 3-port cavity with $\rho_1^2=0.7,\rho_0^2=0.8$,$\eta_2=\eta_{2 \mathrm{max}}$ and $\rho_3^2=0.9$ as a function of the detunings $\Phi_1$ and $\Phi_2$. The maxima of the transmitted power are positioned at the crossings of the individual resonance branches, e.g. at $\Phi_1=78^\circ,\Phi_2=-6.5^\circ$ and $\Phi_1=102^\circ,\Phi_2=6.5^\circ$. At this operating point only about half of the light is transmitted. The lines in the zoomed-in figure on the right represent  $\Phi_1^{\mathrm{res}}=-0.5 \mathrm{arg}[ \rho_{\mathrm{g}_2\mathrm{m}_3}(\Phi_2)]$ (dashed) and  $\Phi_2^{\mathrm{res}}=-0.5 \mathrm{arg}[ \rho_{\mathrm{g}_2\mathrm{m}_1}(\Phi_1)]$ (dotted), respectively.}
 	\label{fig:Fig09}
 \end{figure}

\begin{figure}[htbp]
 	\centering
 		\includegraphics[height=4.5cm]{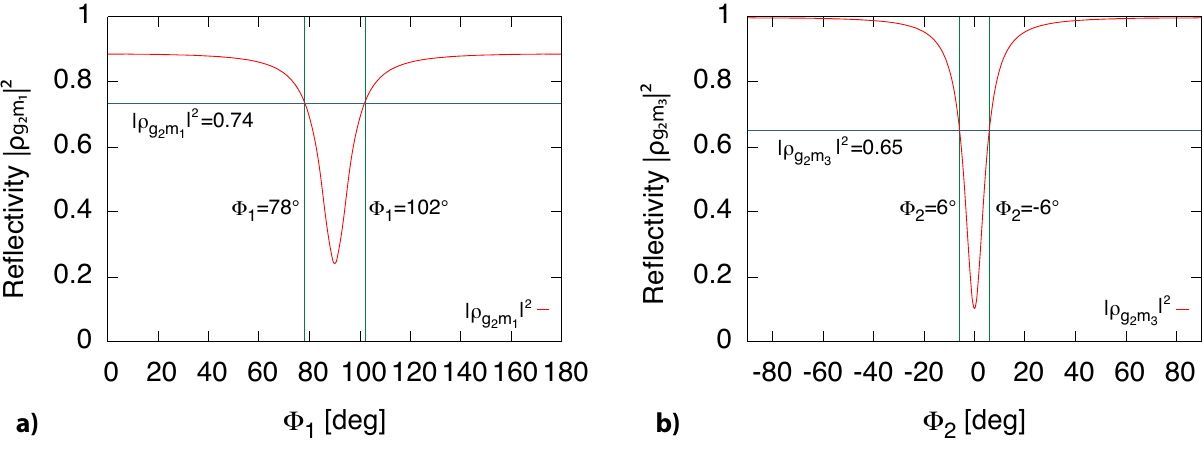}
 	\caption{Compound mirror reflectivities $|\rho_{\mathrm{g}_2\mathrm{m}_1}|^2$ (a) and $|\rho_{\mathrm{g}_2\mathrm{m}_3}|^2$ (b) for the $\eta_{2 \mathrm{max}}$-configuration as a function of $\Phi_1$ and $\Phi_2$, respectively. The maximum of the transmitted power (see Figure \ref{fig:Fig09}, with $\Phi_1=78^\circ$ and $\Phi_1=102^\circ$, respectively, and $\Phi_2=\pm6^\circ$) does not correspond to the detuning where $|\rho_{\mathrm{g}_2\mathrm{m}_3}|^2=\rho_1^2=0.7$.}
 	\label{fig:Fig10}
 \end{figure}

Figure~\ref{fig:Fig09} shows the resonance pattern of a power-recycled 3-port grating cavity with $\rho_1^2=0.7$, $\rho_0^2=0.8$, $\rho_3^2=0.9$, and $\eta_2^2=\eta^2_{2 \mathrm{max}}=(1+\rho_0)^2/4 \approx 0.897$. The resonance pattern is periodic in $\Phi_1\,\mathrm{mod}\,\pi$ and in $\Phi_2\,\mathrm{mod}\,\pi$ and similar to the one of the three-mirror cavity. The maxima are aligned along the resonance branches of the individual cavities given by the detunings required to compensate the phase shifts from the other cavity, respectively. A maximum of the transmitted power is located at $\Phi_1=78^\circ,\Phi_2=-6.5^\circ$ and $\Phi_1=102^\circ,\Phi_2=6.5^\circ$. But even on peak resonance only about half of the light is transmitted through the arm cavity. It can be shown that the major part of the missing power is forward-reflected, whereas almost no power is reflected back towards the laser field input. \\
Also the compound mirror reflectivities show a similar behavior like for a three-mirror cavity (see Fig.~\ref{fig:Fig10}). The reflectivity of the grating cavity is maximal on anti-resonance ($\Phi_2=90^\circ$) and minimal for $\Phi_2=0^\circ$. But here, a maximum of the transmitted power does not correspond to a detuning where $|\rho_{\mathrm{g}_2\mathrm{m}_3}|^2$ matches $\rho_1^2$. For all detunings the compound mirror power reflectivity of the first cavity is lower than $\rho_3^2$. It is interesting to note that  $|\rho_{\mathrm{g}_2\mathrm{m}_3}|^2$ can be close to unity for a large detuning of the arm cavity, which in conjunction with a proper choice of $\rho_1$ will lead to a very large power enhancement inside the power-recycling cavity. However, the power enhancement factor of the arm cavity is very small. {The product of both power enhancement factors can still reach 20,000 being as high as in the conventional case and as in our all-reflective "case 1".} But in a realistic scenario, for example when applied to a Michelson interferometer the high power in the power-recycling cavity produces unnecessarily high optical loss, e.g in terms of scattering. From this we conclude that the case considered here is ill-suited for an application in power-recycled laser interferometry. 

\subsection{Case 3: $\eta_2=\eta_{2 \mathrm{mid}}$} 
\begin{figure}[htbp]
	\centering
		\includegraphics[height=4.05cm]{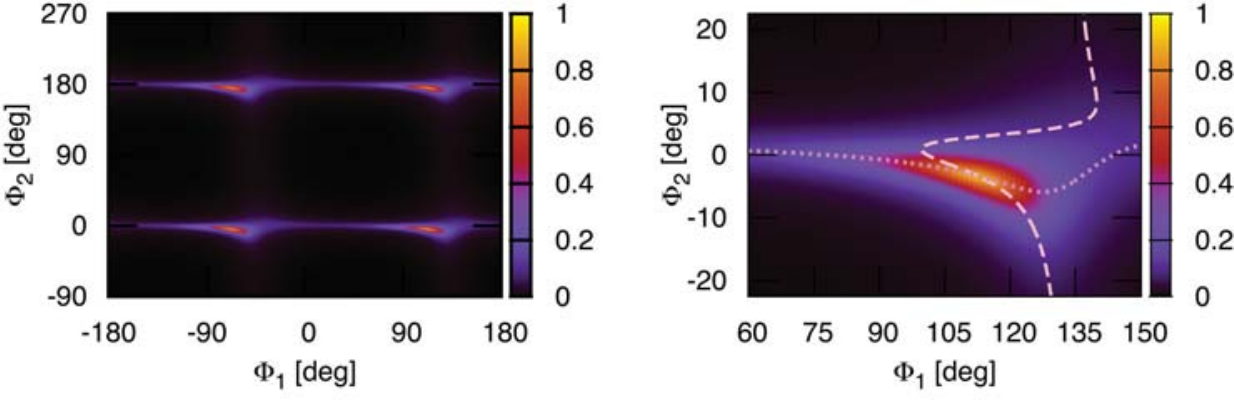}
	\caption{Transmitted power at a power-recycled 3-port cavity with $\rho_1^2=0.7,\rho_0^2=0.8$,$\eta_2^2=\eta_0^2=0.45$ and $\rho_3^2=0.9$ as a function of the detunings $\Phi_1$ and $\Phi_2$. The operating point $\Phi_1=113^\circ,\Phi_2=-3.3^\circ$ corresponds to a maximum of the internal power in the arm cavity. The lines in the zoomed-in figure on the right represent  $\Phi_1^{\mathrm{res}}=-0.5 \mathrm{arg}[ \rho_{\mathrm{g}_2\mathrm{m}_3}(\Phi_2)]$ (dashed) and $\Phi_2^{\mathrm{res}}=-0.5 \mathrm{arg}[ \rho_{\mathrm{g}_2\mathrm{m}_1}(\Phi_1)]$ (dotted), respectively.}
	\label{fig:Fig11}
\end{figure}

\begin{figure}[htbp]
	\centering
		\includegraphics[height=4.50cm]{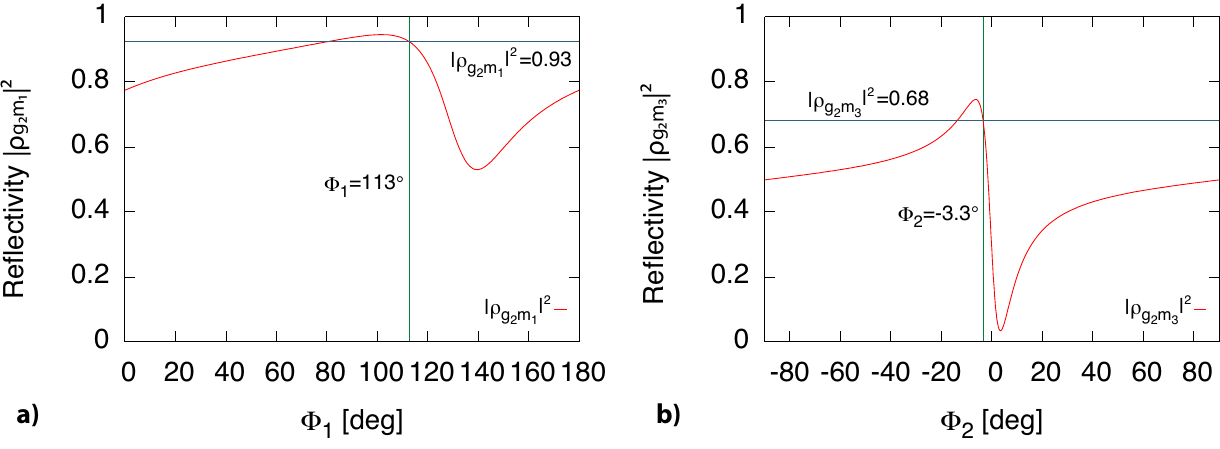}
	\caption{Compound mirror reflectivities $|\rho_{\mathrm{g}_2\mathrm{m}_1}|^2$ (a) and $|\rho_{\mathrm{g}_2\mathrm{m}_3}|^2$ (b) for $\eta_2^2=\eta_0^2$ as a function of $\Phi_1$ and $\Phi_2$, respectively. Note that $|\rho_{\mathrm{g}_2\mathrm{m}_3}|^2$ is not symmetric around $\Phi_1=0^\circ$. As a consequence no resonance doublets occur in Figure \ref{fig:Fig11}. The maximum of the transmitted power does not correspond to the detuning where $|\rho_{\mathrm{g}_2\mathrm{m}_1}|^2=\rho_3^2=0.9$ and $|\rho_{\mathrm{g}_2\mathrm{m}_3}|^2=\rho_1^2=0.7$, respectively.}
	\label{fig:Fig12}
\end{figure}

Fig.~\ref{fig:Fig11} shows the resonance pattern of a power-recycled 3-port grating cavity with $\rho_1^2=0.7$, $\rho_0^2=0.8$, $\rho_3^2=0.9$, and $\eta_2^2=\eta_0^2=0.45$. The pattern does not show the symmetry that is present in the $\eta_{2 \mathrm{max}}$-configuration. This can be understood when looking at the compound mirror power reflectivities shown in Fig.~\ref{fig:Fig12}. In contrast to the previously investigated configurations, $|\rho_{\mathrm{g}_2\mathrm{m}_3}|^2$ is not symmetric around the resonance of the arm cavity at $\Phi_2=0^\circ$. As a consequence no resonance doublets occur as found in the $\eta_{2 \mathrm{max}}$-configuration, because only for a negative detuning the compound mirror power reflectivity is in a regime that allows for significant power enhancement in the recycling cavity. The maximum in Fig.~\ref{fig:Fig11} corresponds to a transmission of {89\%}.\\
Let us now consider again the parameter set that is more relevant for a possible application in power-recycled interferometry. For a cavity with the parameters $\rho_0^2=0.95,\eta_2^2=\eta_0^2=0.475$ and $\rho_3^2=0.99995$ the reflectivity of the recycling mirror has to be chosen to $\rho_1^2=0.999$ in order to obtain impedance matching. Although this reflectivity is significantly higher than in our "case 1", the power-recycling factor remains $\approx 512$. The reason is a slightly different operating point of $\Phi_1=112^\circ,\Phi_2=-0.7^\circ$. The power enhancement factor of the cavity is again 78. Note that this factor is more or less the same as in "case 1", since the value of $\Phi_2=-0.7^\circ$ is very close to the arm cavity resonance. The overall power enhancement therefore is again 20,000. 
It is a somewhat surprising result that a grating coupler with $\eta_2 > \eta_{2\mathrm{min}}$ allows for the same power-enhancement factors as in "case 1". As a result, gratings of our "case 3" are also useful for applications in power-recycled interferometers, similar to gratings of "case 1".

\section{Conclusion} 
\label{sec:conclusion}

We have investigated the all-reflective optical coupling of two cavities. We have shown that all-reflective power-recycling of a Fabry-Perot arm cavity is possible in a rather efficient way using a (low diffraction efficiency) 3-port grating in second-order Littrow configuration. In such a configuration neither of the two arm cavity mirrors transmit laser light. Such a coupling scheme therefore reduces optical absorption in substrate materials. It may also allow for the selection of new substrate materials with higher thermal conductivities (but possibly larger absorption coefficients) to reduce the effect of possible light absorption in the substrate's surface, which might occur if the reflection grating incorporates a high-reflectivity dielectric coating stack. Our coupling scheme may therefore enable the cooling of mirrors down to temperatures not achievable with a transmissive coupling scheme. An important result of our analysis is that high power build-ups in the arm cavities together with a high power-recycling factor due to the upstream cavity are possible for a rather unexpected broad range of grating diffraction efficiencies. In Ref.~\cite{bunkowski1} it was shown that a single 3-port grating coupled Fabry-Perot cavity can only show a high reflectivity if the 3-port grating's second-order diffraction efficiency is close to its minimal value. One could therefore expect that power-recycling of such a Fabry-Perot cavity is only possible if the same condition is fulfilled. But here we have shown that this condition is not mandatory. Even for rather high second-order diffraction efficiencies (our "case 3") the same power build-up factors are possible as for minimal second-order diffraction efficiencies (our "case 1"). This result considerably relaxes the demands on grating fabrication. We found that for all 3-port grating coupled power-recycled arm cavities the laser power inside the recycling cavity is at least twice as high as for the conventionally coupled scheme. Since in ultra-sensitive interferometers only the thermal noise of the arm cavity mirrors needs to be reduced through cryogenic cooling this effect is not crucial.\\
We would like to mention that nano-structured grating surfaces produce relatively large levels of light scattering compared with dielectrically coated super-polished plane surfaces. Scattered photons can limit the sensitivity of interferometric devices operating at signal frequencies in the audio-band \cite{Vahlb07NJP}. However, there are great potentials to improve both systems in future. Scattering in a nano-structure arises from deviations from the perfectly periodic grating structure. It can be expected that  lithography technology will be improved and this problem will be reduced. In gravitational wave detectors scattered light is a problem because some parts \textit{re-}scatter into the interferometer mode. Future detectors might use appropriately designed baffles in order to suppress re-scattering. \\
We conclude that the all-reflective coupling of optical cavities based on 3-port diffraction gratings provide a valuable concept for power-recycled interferometers at low temperatures.

\section*{Acknowledgements} This research was supported by the Deutsche Forschungsgemeinschaft with the Sonderforschungsbereich TR7 and is part of the Centre for Quantum Engineering and Space-Time Research QUEST.


\begin{thebibliography}{99}

\bibitem{DHD87} B. Dahmani, L. Hollberg, and R. Drullinger, "Frequency stabilization of semiconductor lasers by resonant optical feedback," Opt. Lett. 12, 876-878 (1987).

\bibitem{Wicht05} A.~Wicht, P.~Huke, R.-H.~Rinkleff, "Advancing the Optical Feed Back Concept: Grating Enhanced External Cavity Diode Laser," Physica Scripta. {\bf T118}, 82-84 (2005)

\bibitem{BraginskyBar} V.~B.~Braginsky, M.~L.~Gorodetsky, F.~Ya.~Khalili, "Optical bars in gravitational wave antennas," Phys. Lett. A \textbf{232}, 340-348 (1997).

\bibitem{meers}	B.~J.~Meers, "Recycling in laser-interferometric gravitational-wave detectors," Phys. Rev. D {\bf 38}, 2317-2326 (1988).

\bibitem{recycling}	P.~Fritschel, D.~Shoemaker, R.~Weiss, ``Demonstration of light recycling in a Michelson interferometer with Fabry-Perot cavities``, Appl. Opt. {\bf 31}, 1412-1418 (1992).
			 
\bibitem{MSNCSRWD93} J.~Mizuno, K.~A.~Strain, P.~G.~Nelson, J.~M.~Chen, R.~Schilling, A.~R{\"u}diger, W.~Winkler and K.~Danzmann, "Resonant sideband extraction: a new configuration for interferometric gravitational wave detectors," Phys. Lett. A {\bf 175}, 273-276 (1993).

\bibitem{HMSRWD96} G.~Heinzel, J.~Mizuno, R.~Schilling, A.~R{\"u}diger, W.~Winkler, and K.~Danzmann, "An experimental demonstration of resonant sideband extraction for laser-interferometric gravitational wave detectors," Phys. Lett. A {\bf 217}, 305-314 (1996).

\bibitem{mizuno} J.~Mizuno, "Comparison of optical configurations for laser-interferometric gravitational-wave detectors," Internal report MPQ {\bf 203} (1995).

\bibitem{TSLD07} A.~Th\"uring, R.~Schnabel, H.~L\"uck, and K.~Danzmann, "Detuned Twin-Signal-Recycling for ultrahigh-precision interferometers," Opt. Lett. \textbf{32}, 985-987 (2007).

\bibitem{TGVMDS09} A.~Th\"uring, C.~Gr\"af, H.~Vahlbruch, M.~Mehmet, K.~Danzmann, and R.~Schnabel, "Broadband squeezing of quantum noise in a Michelson interferometer with Twin-Signal-Recycling," Opt. Lett. \textbf{34}, 824-826 (2009).

\bibitem{LIGO2009}	B.~P.~Abbott et al., "LIGO: the Laser Interferometer Gravitational-Wave Observatory," Rep. Prog. Phys. {\bf 72}, 076901 (2009).

\bibitem{Virgo2006}	F.~Acernese et al., "The Virgo status," Class.Quant.Grav. {\bf 23}, S635-S642 (2006). 

\bibitem{lensing} K.~A. Strain, K.~Danzmann, J.~Mizuno, P.~G.~Nelson, A.~R\"udiger, R.~Schilling and W.~Winkler,  "Thermal lensing in recycling interferometric gravitational-wave detectors," {Phys. Lett. A} {\bf 194}, 124-132 (1994).
	 
\bibitem{BGV99} V. B. Braginsky , M. L. Gorodetsky and S. P. Vyatchanin, "Thermodynamical fluctuations and photo-thermal shot noise in gravitational wave antennae," Phys. Lett. A \textbf{264}, 1-10 (1999).

\bibitem{BGV00} V.~B.~Braginsky, M.~L.~Gorodetsky, and S.~P.~Vyatchanin, "Thermo-refractive noise in gravitational wave antennae," Phys. Lett. A \textbf{271}, 303-307 (2000).

\bibitem{Cryo09} K. Arai \textit{et al.}, "Status of Japanese gravitational wave detectors," Class. Quantum Grav. 26 204020 (2009).

\bibitem{Levin98} Y.~Levin, "Internal thermal noise in the LIGO test masses: A direct approach," \emph{Phys. Rev. D} {\bf 57}, 659-663 (1998).

\bibitem{Drever} R.~W.~P.~Drever, "Concepts for Extending the Ultimate Sensitivity of Interferometric Gravitational Wave Detectors Using Non-Transmissive Optics with Diffractive or Holographic Coupling," in \textit{Proceedings of the Seventh Marcel Grossman Meeting on General Relativity}, M.~Keiser and R.~T.~Jantzen (eds.), World Scientific, Singapore (1995). 

\bibitem{Sun} K.-X.~Sun and R.~L.~Byer, "All-reflective Michelson, Sagnac, and Fabry-Perot interferometers based on grating beam splitters," Opt. Lett.  {\bf 23}, 567-569 (1998). 

\bibitem{bunkowski2} A.~Bunkowski, O.~Burmeister, P.~Beyersdorf, K.~Danzmann, R.~Schnabel, T.~Clausnitzer, E.-B.~Kley, A.~T\"unnermann "Low-loss grating for coupling to a high-finesse cavity," {Opt. Lett.} {\bf 29}, 2342-2344 (2004).

\bibitem{higheff1} A.~Bunkowski, O.~Burmeister, K.~Danzmann, R.~Schnabel, T.~Clausnitzer, E.-B.~Kley, A.~T\"unnermann, "Optical characterization of ultra-high diffraction efficiency gratings," {Appl. Opt} {\bf 45}, 5795-5799 (2006).
	
\bibitem{gratingbs} D.~Friedrich, O.~Burmeister, A.~Bunkowski, T.~Clausnitzer, S.~Fahr, E.-B.~Kley, A.~T\"unnermann, K.~Danzmann, R.~Schnabel, "Diffractive beam splitter characterization via a power-recycled interferometer," {Opt. Lett.} {\bf 33}, 101-103 (2008).

\bibitem{BruecknerPRL} F. Br\"uckner, D. Friedrich, T. Clausnitzer, M. Britzger, O. Burmeister, K. Danzmann, E.-B. Kley, A. T\"unnermann, and R. Schnabel, "Realization of a monolithic high-reflectivity cavity mirror from a single silicon crystal," Phys. Rev. Lett., accepted (2010).
	
\bibitem{bunkowski1} A.~Bunkowski, O.~Burmeister, K.~Danzmann, and R.~Schnabel, "Input-output relations for a 3-port grating coupled Fabry-Perot cavity," {Opt. Lett.} {\bf 30}, 1183-1185 (2005).

\bibitem{Vahlb07NJP} H.~Vahlbruch, S.~Chelkowski, K.~Danzmann, and R.~Schnabel, "Quantum engineering of squeezed states for quantum communication and metrology," New J. Phys. \textbf{9}, 371 (2007).


\end{thebibliography}
\end{document}